\documentclass[superscriptaddress,aps,pra,twocolumn,showkeys]{revtex4-2}
\usepackage{amsmath,mathtools,siunitx}
\usepackage{amssymb}
\usepackage{graphicx}
\usepackage{color}
\usepackage{bm}
\usepackage[colorlinks,linkcolor=blue,anchorcolor=blue,citecolor=blue,urlcolor=blue]{hyperref}
\usepackage{booktabs}
\usepackage{braket} 
\usepackage{bm}
\usepackage{subcaption}
\usepackage{caption}

\begin{document}
\title{Quantum phase transition in two dimension nonlinear cavity optomagnonic system}

\author{Yu Sang}
\affiliation{Lanzhou Center for Theoretical Physics, Key Laboratory of Theoretical Physics of Gansu Province, and Key Laboratory of Quantum Theory and Applications of MoE, Lanzhou University, Lanzhou, Gansu 730000, China}

\author{Jie Liu}
\affiliation{Lanzhou Center for Theoretical Physics, Key Laboratory of Theoretical Physics of Gansu Province, and Key Laboratory of Quantum Theory and Applications of MoE, Lanzhou University, Lanzhou, Gansu 730000, China}

\author{Lei Tan}
\email{tanlei@lzu.edu.cn}
\affiliation{Lanzhou Center for Theoretical Physics, Key Laboratory of Theoretical Physics of Gansu Province, and Key Laboratory of Quantum Theory and Applications of MoE, Lanzhou University, Lanzhou, Gansu 730000, China}

\begin{abstract}
The superfluid-Mott insulator and ergodic-many body localization transitions based on a two dimension nonlinear cavity optomagnonic system are investigated, where a yttrium iron garnet (YIG) sphere is embedded at each site of a two-dimensional coupled cavity array. It can be demonstrated that, the introduction of phonon–photon coupling enhances the coherence of the system when considering the Kerr nonlinearity of the YIG sphere. In contrast, the Kerr nonlinearity of photons is more conducive to the Mott insulating phase than that of the YIG sphere. We further elucidate the underlying physical mechanism by calculating the effective repulsive potential of the system in the presence of photonic Kerr nonlinearity. Regarding the ergodic–many body localization transition, the results indicate that as the disorder strength of the Kerr nonlinearity increases, the system transitions from the ergodic phase to the many body localized phase, while increasing the chemical potential expands the region of the ergodic phase. This work provides a novel framework for characterizing quantum phase transitions in cavity optomagnonic systems and offers an experimentally feasible scheme for studying them, thereby yielding valuable insights for quantum simulation.

\end{abstract}

\maketitle
\section{INTRODUCTION}

Quantum simulation provides a highly practical platform for investigating complex quantum many-body phenomena such as quantum topology~\cite{4, 5, 15}, quantum phase transitions~\cite{6, 7, 8}, and quantum thermalization~\cite{1, 2, 3}. Coupled cavity array systems based on quantum optics and their extensions, as models of light-matter interaction, have been extensively used to study quantum many-body problems~\cite{9, 10, 11, 45, 13}. Under the combined effects of mechanisms such as photon tunneling, light-matter coupling interactions~\cite{16, 9}, and disorder perturbations~\cite{18, 19}, these models exhibit complex and diverse quantum effects. Among them, the Jaynes-Cummings Hubbard (JCH) model, which involves embedding a two-level atom in each resonant cavity, has been widely applied in recent years~\cite{19, 20, 21, 22, 23, 24, 25, 26, 27, 28}. In the JCH model, by tuning the light-atom interaction and photon tunneling, superfluid–Mott insulator phase transitions and ergodic–many body localization (MBL) phase transitions can be realized~\cite{16, 9}. The superfluid–Mott insulator phase transition arises from the competition between the system's repulsive potential and photon tunneling.  On the other hand, the ergodic–MBL phase transition results from disorder effects in the JCH model~\cite{18, 19}. Due to the randomness of atomic positions within the cavities, such disorder effects can be introduced through atom-light interactions~\cite{19}. (It is worth noting that whether the MBL transition constitutes a true phase transition or a crossover remains debated~\cite{29, 30, 31, 32}; this paper treats it as a phase transition~\cite{32} and does not engage in this debate.)

In recent years, the cavity optomagnonic system combining different degrees of freedom, such as photonic, mechanical, electronic, or magnetic has triggered immense interests in recent years, in which a strong coherent coupling between microwave photons and magnons in Yttrium Iron Garnet (YIG) was demonstrated. 
Due to their advantages of high reliability, low loss, high Curie temperature 
and long coherence times, a large number of extraordinary quantum effects have been studied, including: nonreciprocal quantum phase transition~\cite{58}, Quantum magnon conversion~\cite{59}, the high-order sideband generation~\cite{65}, the self-sustained oscillations and chaos~\cite{66,67}, entanglement~\cite{68,69,70,71,72}, non-Hermitian physics~\cite{73, 74}, coherent and dissipative magnon-photon
interaction~\cite{75,76,77,78}, magnon squeezing~\cite{79}, magnon Fock
state~\cite{80}, ultrastrong-coupling induced photon squeezing~\cite{60}.
The cavity optomagnonic system has emerged as a promising platform for quantum simulation. Our group has previously explored this direction by introducing YIG spheres into the JCH model to enhance the degrees of freedom and study their influence on quantum phase transitions~\cite{9}, as well as investigating phase transitions in an open cavity optomagnonic system containing a YIG sphere~\cite{61}.

Thus, it is interesting to make a thorough inquiry how the extraordinary characteristics of the hybrid cavity-magnonics system can be harnessed for the purpose 
of quantum simulations. Inspired by these improvements, a unified physical model by replacing the two-level atom in the JCH model with YIG are constructed to 
investigate two types of quantum phase transitions, i.e., the superfluid-Mott insulator and the ergodic-MBL transition. Based on the 
mean-field methods and the exact diagonalization approach, the effects of Kerr nonlinearity, the cavity-magnonics coupling strength other controlling parameters are analyzed. To the best of our knowledge, no similar work has been reported.

It can be found that, the photon Kerr nonlinearity strongly favors the localized Mott state compared to the Kerr nonlinearity of the YIG sphere. Moreover, when introducing optomechanical coupling into the cavity
optomagnonic system with Kerr nonlinearity from the YIG sphere, the regions of superfluid state significantly expands.
On the other hand, the transitions from the ergodic phase to the MBL phase can be observed as disorder increases. Furthermore, the increasing of the chemical potential leads to an expansion of the ergodic phase region. Our research may open up a wealth of possibilities for deep understanding the mechanism of the quantum phase transiton and experimentally.

The structure of this paper is organized as follows. In Sec. \ref{II}, we provide a detailed introduction to the cavity optomagnonic array system that serves as the platform for investigating quantum phase transitions. In Sec. \ref{III}, we present an in-depth analysis of the superfluid–Mott insulator phase transition, highlighting the intricate interplay between interactions and coherence. The discussion then turns to the ergodic–many body localization phase transition in Sec. \ref{IV}, where the effects of disorder and interactions are systematically examined. Finally, a comprehensive summary of our findings is given in Sec. \ref{V}.

\section{Physical models and Hamiltonian}\label{II}

\begin{figure}[!htpb]
	\begin{subfigure}[b]{0.3\textwidth} 
		\includegraphics[width=\linewidth]{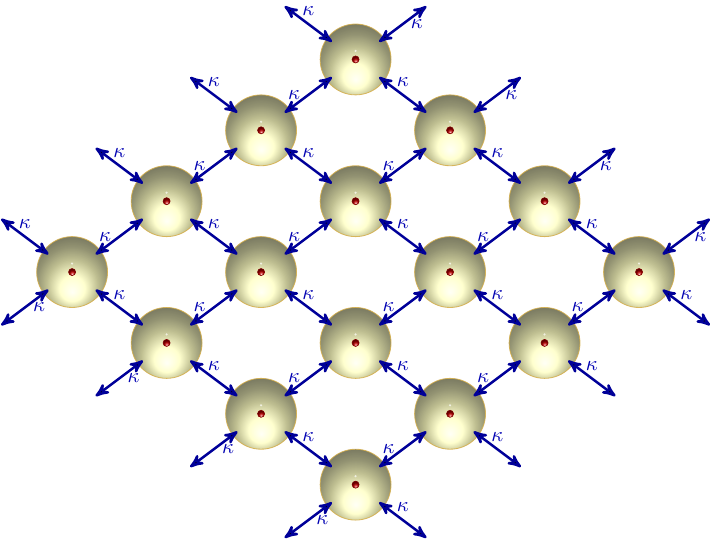}
		\label{fig:sub1.1}
	\end{subfigure}
	\hfill 
	\begin{subfigure}[b]{0.15\textwidth} 
		\includegraphics[width=\linewidth]{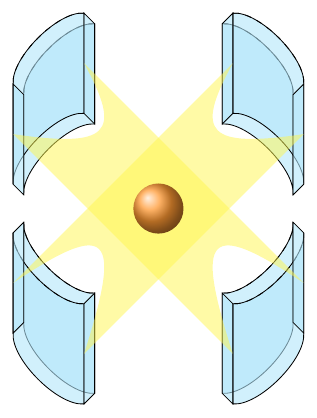} 
		\label{fig:sub1.2}
	\end{subfigure}
	\captionsetup{justification=raggedright, singlelinecheck=false}
	\caption{(a) Schematic diagram of a two-dimension-\\al coupled optomagnonic cavity array. Each yellow sphere represents\quad an\quad optical cavity containing a \\ YIG sphere. The photon hopping strength between neighboring cavities is $\kappa$. (b) Schematic diagram of\\ a single cavity embedding a YIG sphere.}
	\label{fig:figure1}
\end{figure}

We consider the model illustrated in Fig.~\ref{fig:figure1}, where each cavity contains a YIG sphere interacting with the photon mode (see Fig.~\ref{fig:figure1}). The nearest neighboring cavities are connected via photon hopping. The Hamiltonian of the system can be written as:

\begin{align}
	H^{cop}_{i} =& \left( \omega_c a_i^{\dagger} a_i + \omega_m m_i^{\dagger} m_i \right)
	+ \sum_{i} \left( g_{1} m_{i}^{\dagger} a_{i} + h.c. \right) \notag \\
	&  
	 - \mu \sum_{i} N_{i} 
	 \label{H_{cop}}
\end{align}

\begin{align}	
	H^{system} = & \sum_{i} H^{cop} - \kappa \sum_{\langle i,j \rangle} a^{\dagger}_{i} a_{j} - g_{m} \sum\limits_{i} a_{i}^{\dagger}a_{i}(b^{\dagger}_{i}+b_{i})\notag \\
	& + K \sum_{i} m^{\dagger}_i m_{i} m^{\dagger}_i m_{i}
	 + U\sum\limits_{i}a^{\dagger}_{i} a^{\dagger}_{i} a_{i} a_{i}
	\label{H_{system}}
\end{align}

In Eq. (\ref{H_{cop}}), the first term describes the free Hamiltonian of photons and magnons in each cavity, where $\omega_c$ and $\omega_m$ denote the photon and magnon frequencies, respectively. Here, $a^{\dagger}_i$ ($a_i$) and $m^{\dagger}_i$ ($m_i$) are the creation (annihilation) operators for photons and magnons, satisfying the commutation relations $[a_i, a_j^\dagger] = \delta_{ij}$ and $[m_i, m_j^\dagger] = \delta_{ij}$. The indices $i$ and $j$ label different optomagnonic cavities in the array. The second term describes the photon–magnon coupling within each cavity, where $g_{1}$ is the coupling coefficient. The third part corresponds to the chemical potential, and  $N_i = a_i^{\dagger} a_i + m_i^{\dagger} m_i$ is the total number of excitations of photons and magnons. In Eq. (\ref{H_{system}}), the additional terms beyond $H^{cop}$ are sequentially introduced: a nearest-neighbor tunneling term, with $\kappa$ being the tunneling strength, a phonon-photon coupling term with strength $g_m$, where $b^{\dagger}_{i}$ ($b_{i}$) is the creation (annihilation) operator for phonons; a Kerr nonlinearity term for the magnons with coefficient $K$~\cite{9}; and a Kerr nonlinearity term for the photons with strength $U$~\cite{16}.

To put it simple, a mean-field theory is usually introduced a so called order parameter to provide a measure to investigate the second-order phase transition.
Motivated by Ref.~\cite{20}, we introduce the order parameter $\psi \equiv \langle a_i \rangle$, which is the expectation value of the photon annihilation operator. Since the average electric field is real, we follow Ref.~\cite{8} and take the superfluid order parameter $\psi$ to be real. A vanishing order parameter ($\psi=0$) corresponds to the Mott insulating phase, while $\psi \neq 0$ indicates the superfluid phase. With the mean field assumption, we take decoupling approximation to  investigate our system, i.e. $a^{\dagger}_{i}a_{j} \approx \psi^{\ast}a_{j}+\psi a_{i}^{\dagger} -|\psi|^2$~\cite{20}. The resulting mean-field Hamiltonian, obtained by applying the decoupling approximation to Eq. (\ref{H_{system}}), takes the form of Eq. (\ref{eq:total_hamiltonian}) and is decomposable into a sum of single-site terms, where $z=4$ is the number of the nearest neighbours.

\begin{align}
	H^{PPM} = & \sum_{i}\biggl\{ H^{cop}_{i} - \left[\kappa z \psi \left( a_i + a_i^\dagger \right) - z \kappa |\psi|^2\right] \notag \\
	& - g_{m} a_{i}^{\dagger}a_{i}(b^{\dagger}_{i}+b_{i}) + K \sum_{i} m^{\dagger}_i m_{i} m^{\dagger}_i m_{i} \notag \\
	& + U\sum\limits_{i}a^{\dagger}_{i} a^{\dagger}_{i} a_{i} a_{i} \biggl\}
	\label{eq:total_hamiltonian}
\end{align}

In the absence of phonons, we then construct the basis states $|n, m\rangle$, where $n$ and $m$ denote the number of photons and magnons, respectively, in a given cavity.\quad For a fixed total excitation number $N$, the complete basis is 
\begin{widetext}
$\{|N,0\rangle, |N-1,1\rangle, |N-2,2\rangle \}$. Using this basis,   the Hamiltonian matrix can be explicitly written as, 
\begin{align}
	H^{PM} &=z\kappa \psi^{2}I+ H^{PM} +H^{hop} + {H^{hopT}}\notag\\
	&= \begin{bmatrix}
		H_{(0)}^{PM} & H_{(0)}^{hop} & 0 & 0 & \cdots & \cdots & 0 \\
		H_{(0)}^{hopT} & H_{(1)}^{PM} & H_{(1)}^{hop} & 0 & \cdots & \cdots & \vdots \\
		0 & H_{(1)}^{hopT} & H_{(2)}^{PM} & H_{(2)}^{hop} & \ddots & & \vdots \\
		0 & 0 & H_{(2)}^{hopT} & H_{(3)}^{PM} & \ddots & \ddots & \vdots \\
		\vdots & \vdots & \ddots & \ddots & \ddots &  \ddots & 0 \\
		\vdots & \vdots & & \ddots & \ddots  &  \ddots & \vdots \\
		0 & \cdots & \cdots & \cdots & \cdots & \cdots & H_{(n)}^{PM}
	\end{bmatrix}+z\kappa \psi^{2}I,
\end{align}

where $I$ is the identity matrix and other block matrices $(n>1)$ : $H_{(0)}^{PM} = \left[ 0 \right]$,
$ H_{(0)}^{hop} = \left[-\kappa z \psi \quad 0 \right], $ 

\begin{align}
	H_{(1)}^{PM}=\begin{bmatrix}
		\omega_{c}-\mu & g_{1}\\
		g_{1} & \omega_{m}-\mu+K
	\end{bmatrix},
\end{align}

\begin{align}
	H_{(1)}^{hop}=\begin{bmatrix}
		-\sqrt{2}z\kappa\psi & 0 & 0\\
		0 & -zk\psi & 0\\
		0 & 0 & 0
	\end{bmatrix},
\end{align}

\begin{align}
	&H_{(n)}^{PM}=\notag \\
	&\scalebox{1}{$\begin{bmatrix}
			n\omega_{c}-n\mu+n(n-1)U & \sqrt{n}g_{1} & 0\\
			\sqrt{n}g_{1} & (n-1)\omega_{c}+\omega_{m}-n\mu+K+(n-1)(n-2)U & \sqrt{2(n-1)}g_{1}\\
			0 & \sqrt{2(n-1)}g_{1} & (n-2)\omega_{c}+2\omega_{m}-n\mu+4K+(n-2)(n-3)U
			\label{PM}
		\end{bmatrix}$},
\end{align}
\end{widetext}

\begin{align}
	H_{(n)}^{hop}=\begin{bmatrix}
		-\sqrt{n+1}z\kappa\psi & 0 & 0\\
		0 & -\sqrt{n}\kappa\psi & 0\\
		0 & 0 & -\sqrt{n-1}\kappa\psi
	\end{bmatrix},
	\label{hop}
\end{align}

If the photon-phonon coupling is considered, the basis state is $|n, m, p\rangle$; where $p$ is the phonon number. For simplicity, this study is restricted to the single-phonon case, where the Hamiltonian matrix can be built using the method described above. In this work, we consider only the resonant case $\omega_c = \omega_m$.

This framework enables us to systematically study the superfluid–Mott insulator transition in the two-dimensional cavity–optomagnonic model and explore the effects of the YIG sphere Kerr nonlinearity, photon Kerr nonlinearity, and phonon–photon coupling.

\section{Superfluid–Mott Insulator Phase Transition}\label{III}

\begin{figure}[!htbp]
	\includegraphics[width=0.40\textwidth]{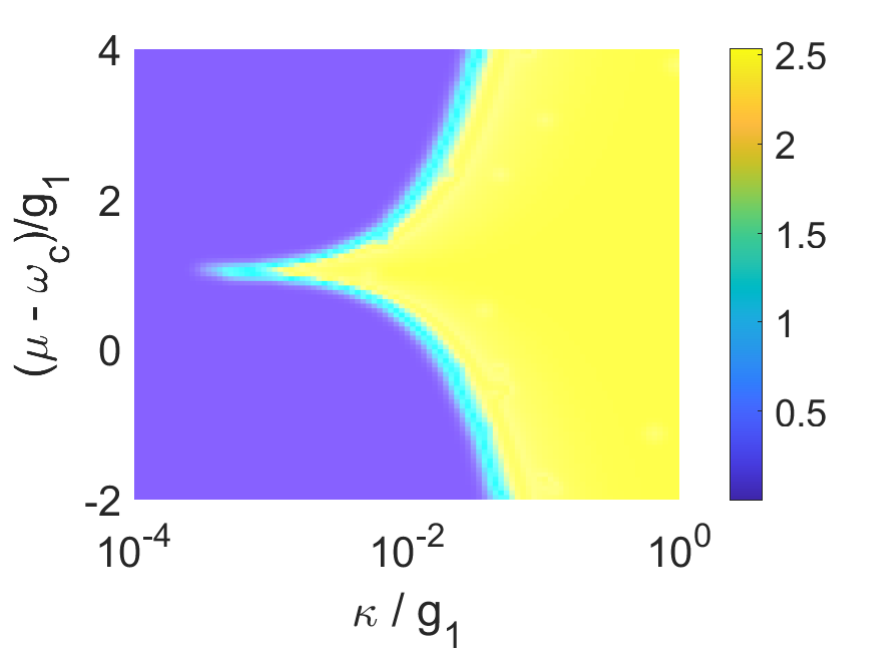}\hfill
	\captionsetup{justification=raggedright, singlelinecheck=false}
	\caption{The phase diagram of the order parameter\\ $\psi$ as a function of the hopping rate $\kappa$ and the rela-\\tive chemical potential $\mu$, plotted for the case whe-\\re $K = U = g_{m} = 0$. Yellow represents the superfl-\\uid phase; blue represents the Mott insulating pha-\\se.  }
	\label{fig:figure2}
\end{figure}

We now turn to a numerical investigation of the superfluid--Mott insulator phase transition. Following the parameter choices of similar systems~\cite{9,16,62}, we first tune the parameters in Hamiltonian~\eqref{eq:total_hamiltonian} such that $K = U = g_{m} = 0$. By applying mean-field theory, we obtain the phase diagram of the superfluid order parameter as a function of the hopping rate and the relative chemical potential, as shown in Fig.~\ref{fig:figure2}. It is clear that the phase diagram separates into two regions according to the value of the order parameter: the superfluid phase and the Mott insulator phase. A notable feature of the phase diagram is the Mott lobes. The Mott insulating phase forms certain lobe structures, while the superfluid phase, which describes the lowest eigenstates excitations, emerges at the edges of the Mott lobes~\cite{16}. As the hopping rate increases, the system undergoes a transition from the localized Mott insulating phase to a coherent superfluid state.

\begin{figure}[!htbp]
	\centering
	\begin{subfigure}[b]{0.23\textwidth}
		\includegraphics[width=\textwidth]{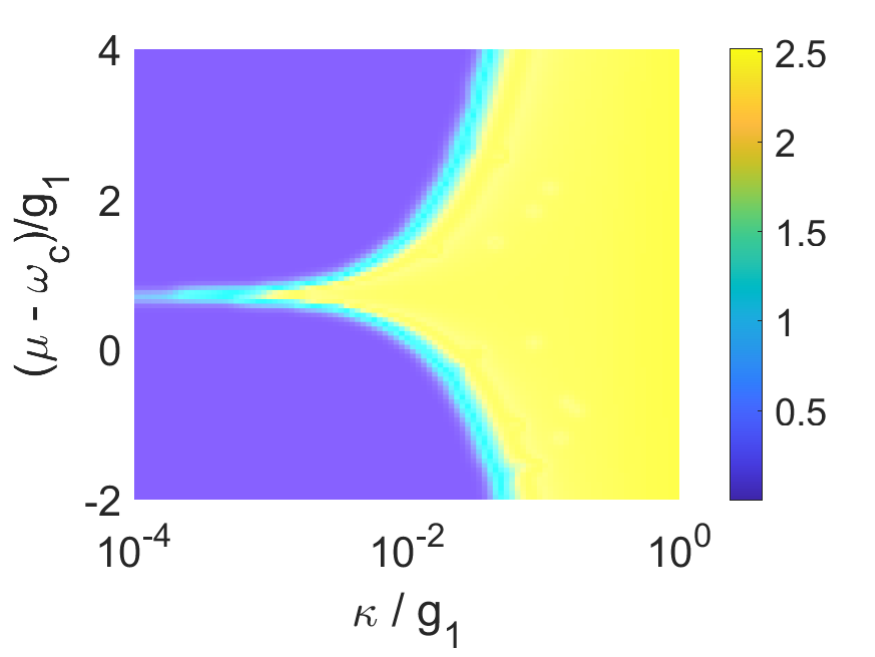}
		\label{fig:yigsub1}
	\end{subfigure}
	\hspace{0.01\textwidth}
	\begin{subfigure}[b]{0.23\textwidth}
		\includegraphics[width=\textwidth]{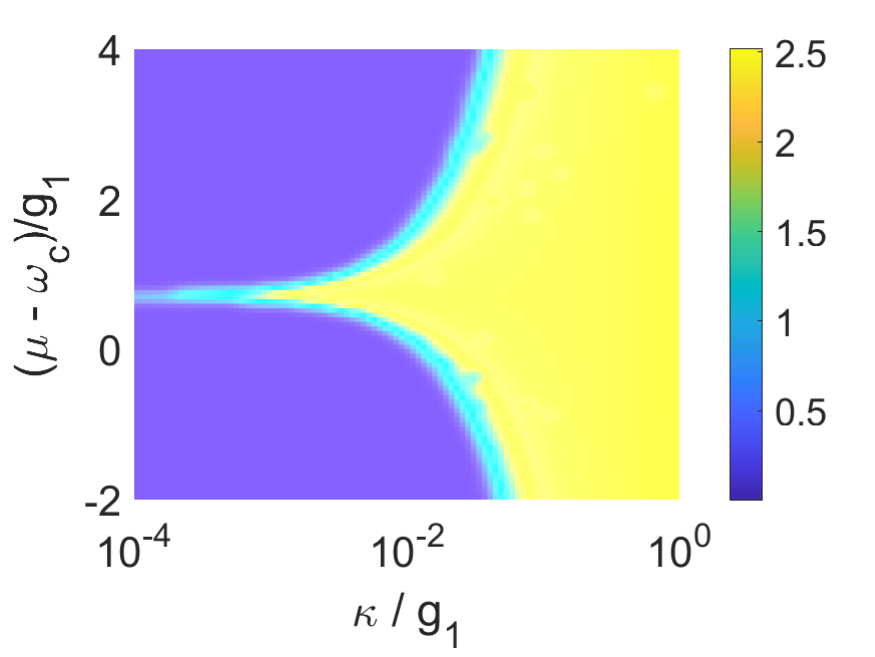}
		\label{fig:yigsub2}
	\end{subfigure}
	\\
	\begin{subfigure}[b]{0.23\textwidth}
		\includegraphics[width=\textwidth]{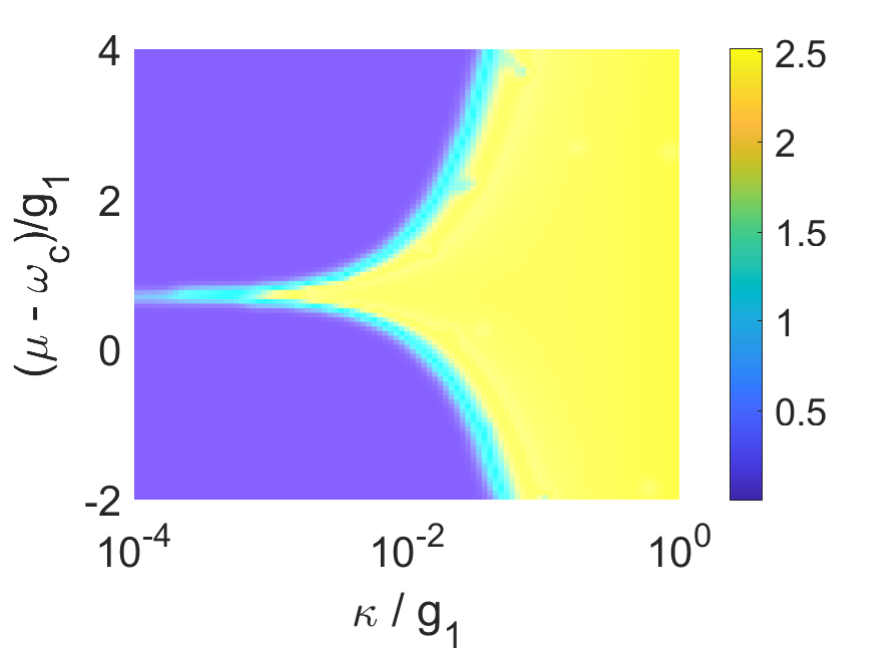}
		\label{fig:yigsub3}
	\end{subfigure}
	\hspace{0.01\textwidth}
	\begin{subfigure}[b]{0.23\textwidth}
		\includegraphics[width=\textwidth]{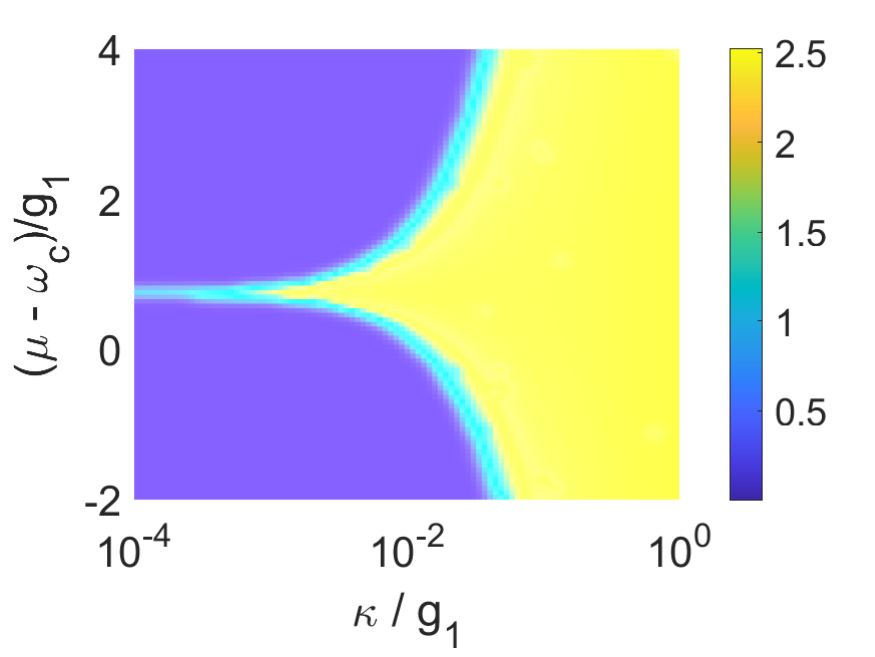}
		\label{fig:yigsub4}
	\end{subfigure}
	\captionsetup{justification=raggedright, singlelinecheck=false}
	\caption{Phase diagram of the order parameter $\psi$ as a function of the hopping rate $\kappa$ and the relative chemi-\\cal potential $\mu$ for different values of $\frac{K}{g_{1}}$. Yellow repre-\\sents the superfluid phase;\quad blue represents the Mott insulating phase. (a) $\frac{K}{g_{1}}=0.001$ \quad(b) $\frac{K}{g_{1}}=0.01$ \quad(c) $\frac{K}{g_{1}}=0.1$ (d) $\frac{K}{g_{1}}= 1$ }
	\label{fig:figure3}
\end{figure}

\begin{figure}[!htbp]
	\centering
	\begin{subfigure}[b]{0.23\textwidth}
		\includegraphics[width=\textwidth]{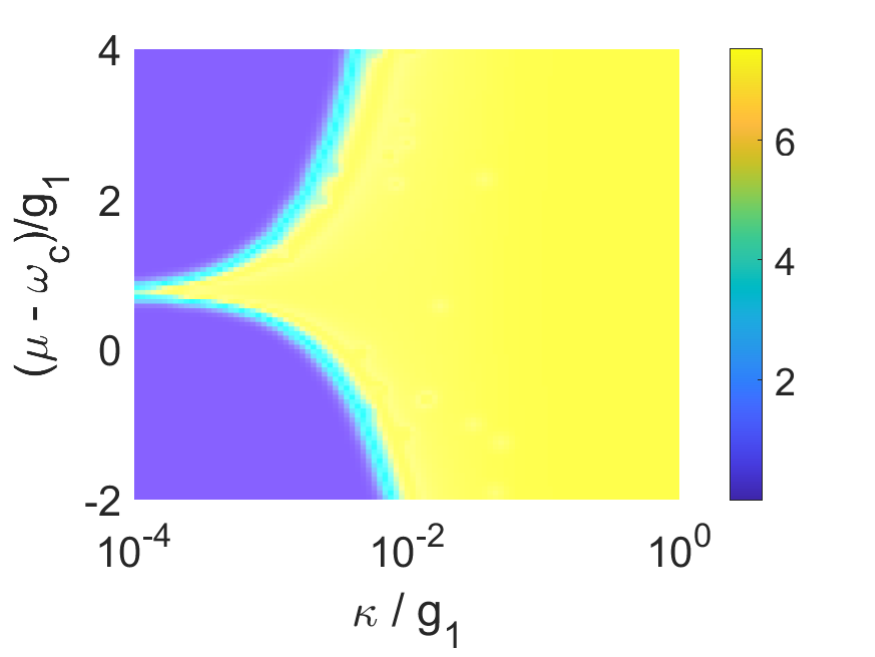}
		\caption{}
		\label{fig:sub1}
	\end{subfigure}
	\hspace{0.01\textwidth}
	\begin{subfigure}[b]{0.23\textwidth}
		\includegraphics[width=\textwidth]{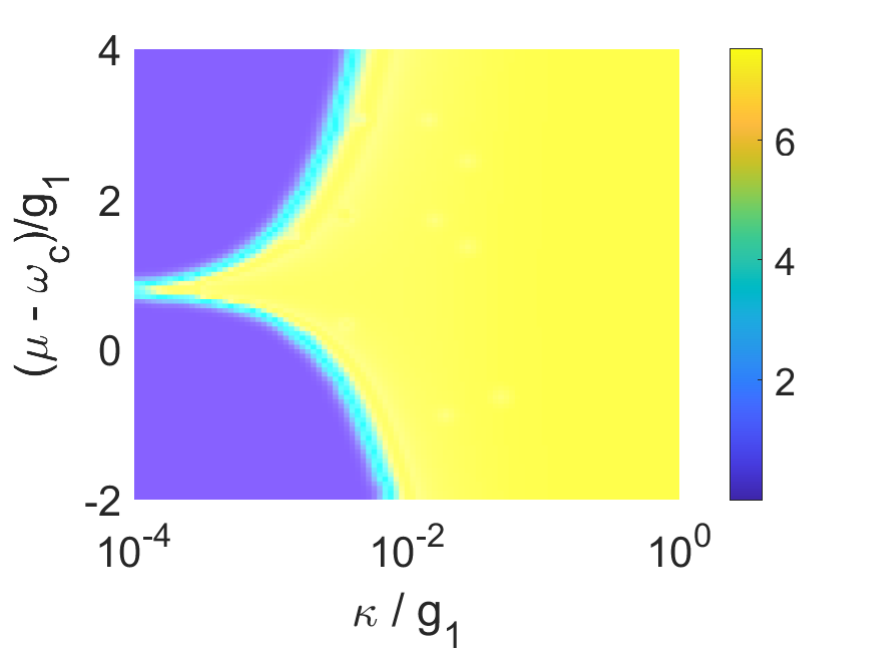}
		\caption{}
		\label{fig:sub2}
	\end{subfigure}
	\captionsetup{justification=raggedright, singlelinecheck=false}
	\caption{Phase diagram of the order parameter $\psi$ as a function of the hopping rate $\kappa$ and the relative chemi-\\cal potential $\mu$ for different values of $\frac{g_{m}}{g_{1}}$. Yellow repr-\\esents the superfluid phase; blue represents the Mott insulating phase. (a) $\frac{g_{m}}{g_{1}}=0.1$ (b) $\frac{g_{m}}{g_{1}}=1$ }
	\label{fig:figure4}
\end{figure}

Next, we consider the Kerr nonlinearity of YIG spheres. In this case, we set $K \neq 0$ while keeping $U = g_{m} = 0$ in Hamiltonian~\eqref{eq:total_hamiltonian}. By varying $K$, we tune the strength of the Kerr nonlinearity of the YIG spheres, and the corresponding phase diagrams are presented in Figs.~\ref{fig:figure3} (a)--(d). Comparing Fig.~\ref{fig:figure2} with Fig.~\ref{fig:figure3}, it can be seen that the regions occupied by the two phases change significantly. Fig.~\ref{fig:figure3} shows that at specific values of the relative chemical potential, the system can still enter the superfluid phase even when the hopping strength is relatively small. This indicates that the hopping effect has a greater impact on the system, and correspondingly, the superfluid region becomes significantly larger. Interestingly, even with a Kerr nonlinearity in Fig.~\ref{fig:figure3} (d) that is a thousand times larger than the one in Fig.~\ref{fig:figure3} (a), the phase diagrams do not exhibit significant changes. This suggests that the presence of Kerr nonlinearity in the YIG sphere influences the coherence of the system to some extent, and the Mott insulating phase remains relatively stable in the presence of Kerr nonlinearity. Subsequently, considering the mobility of the optical cavity and the influence of radiation pressure, it becomes essential to introduce the phonon degree of freedom~\cite{16}. We thus set $K \neq 0$, $g_{m} \neq 0$, and $U = 0$ in Hamiltonian~\eqref{eq:total_hamiltonian}. By tuning $g_{m}$, we obtain the phase diagram shown in Fig.~\ref{fig:figure4}. Compared to Fig.~\ref{fig:figure3}, the area of the superfluid phase further expands, indicating a stronger influence of the hopping effect to the system.

We now briefly discuss the influence of the photon Kerr nonlinearity on the phase diagram. The Kerr nonlinearity of photons has been observed in ultrahigh-quality silica toroidal microcavities~\cite{81}, which can significantly alter the system's behavior, such as the Rabi oscillations in the Jaynes-Cummings model~\cite{82}. Photon-number quantum nondemolition measurements enabled by the optical Kerr effect exhibit extremely high sensitivity in ultrahigh-Q microcavity systems~\cite{63}, and photon ``molecule'' systems formed by two coupled resonators driven into resonance can display pronounced photon antibunching even under very weak Kerr nonlinearity ~\cite{64,16}. Thus, it is of interest to study the effect of such nonlinearities on the quantum phase transitions of cavity-magnon arrays.

\begin{figure}[!htbp]
	\centering
	\begin{subfigure}[b]{0.23\textwidth}
		\includegraphics[width=\textwidth]{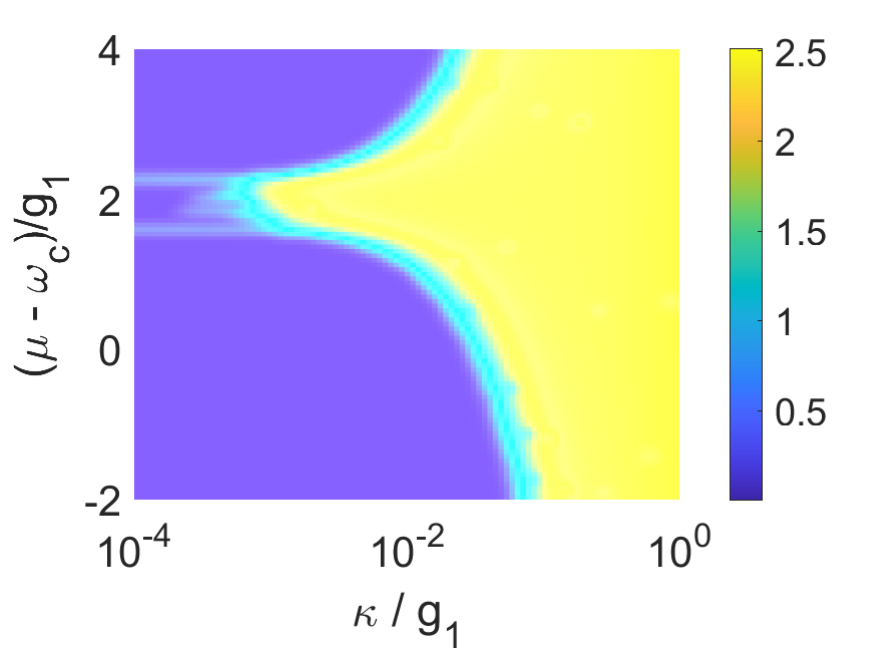}
		\caption{}
		\label{fig:sub1}
	\end{subfigure}
	\hspace{0.01\textwidth}
	\begin{subfigure}[b]{0.23\textwidth}
		\includegraphics[width=\textwidth]{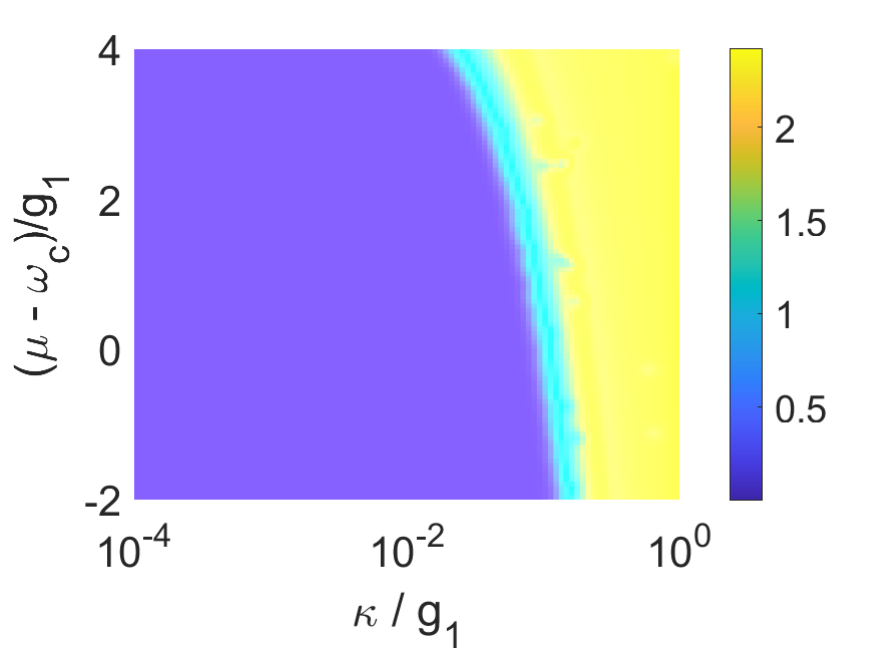}
		\caption{}
		\label{fig:sub2}
	\end{subfigure}
	\\
	\begin{subfigure}[b]{0.23\textwidth}
		\includegraphics[width=\textwidth]{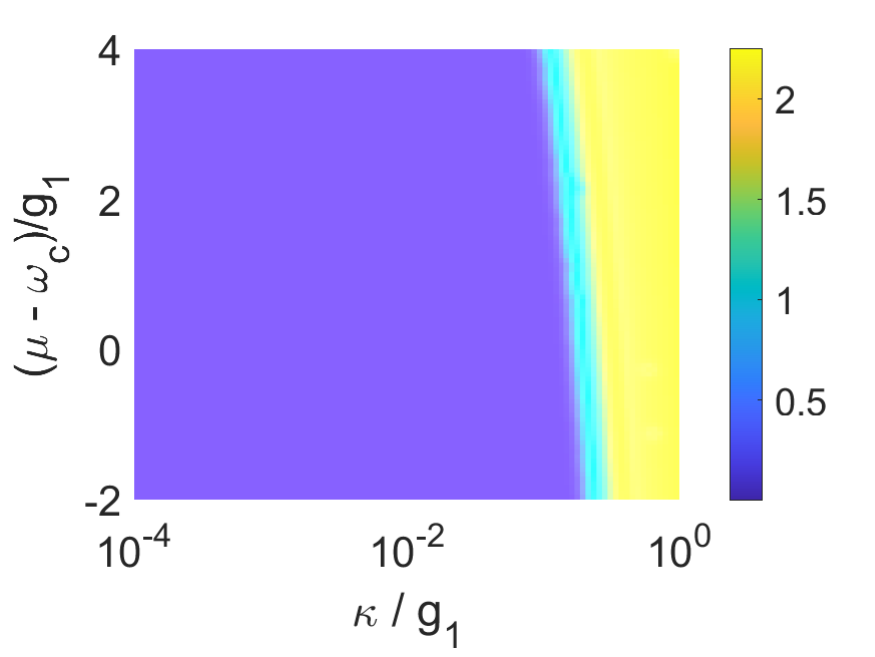}
		\caption{}
		\label{fig:sub3}
	\end{subfigure}
	\captionsetup{justification=raggedright, singlelinecheck=false}
	\caption{Phase diagram of the order parameter $\psi$ as a function of the hopping rate $\kappa$ and the relative chemi-\\cal potential $\mu$ for different values of $\frac{U}{g_{1}}$. Yellow repre-\\sents the superfluid phase;\quad blue represents the Mott insulating phase. (a) $\frac{U}{g_{1}}=0.1$ (b) $\frac{U}{g_{1}}=0.5$ (c) $\frac{U}{g_{1}}=1$ }
	\label{fig:figure5}
\end{figure}

We achieve this by setting $K = 0$, $g_{m} = 0$, and $U \neq 0$ in Hamiltonian~\eqref{eq:total_hamiltonian}. By varying $U$, we control the strength of the photon Kerr nonlinearity and the corresponding phase diagrams is shown in Fig.~\ref{fig:figure5}. It is found that within the calculated parameter range,  as the photon Kerr nonlinearity increases, the Mott lobes gradually disappear and the Mott insulating region expands. To explain this phenomenon, we plot the effective repulsive potential as a function of the photon Kerr nonlinearity.

\begin{figure}[!htbp]
	\centering
	\includegraphics[width=0.35\textwidth]{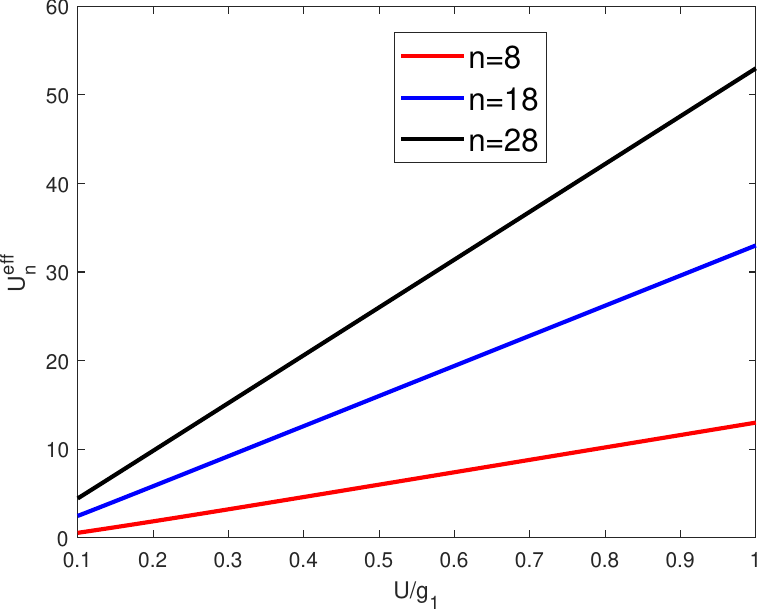}
	\captionsetup{justification=raggedright, singlelinecheck=false}
	\caption{Effective repulsive potential $U^{eff}_{n}$ $(n = 8, 18, 28)$ versus $\frac{U}{g_															{1}}$}
	\label{fig:eff}
\end{figure}

In general, the superfluid--Mott insulator transition can be understood as a competition between the effective on-site repulsive potential and photon hopping. When the effective repulsive potential dominates over the hopping rate, the system resides in the Mott insulating phase; conversely, when hopping dominates, the system enters the superfluid phase. Following the definition in Ref.~\cite{20}, the effective on-site repulsive potential is given by
$U^{\mathrm{eff}}_n = E_{|-,n+1\rangle} - E_{|-,n\rangle} - \omega_{c}$,
where $E_{|-,n\rangle}$ is the lowest eigenenergy of a single cavity with $n$ excitations. As shown in Fig.~\ref{fig:eff}, $U^{\mathrm{eff}}$ increases with $\tfrac{U}{g_{1}}$ within the calculated region. Consequently, the Mott insulating region also expands, consistent with the phase diagrams in Fig.~\ref{fig:figure5}.

From the above analysis, we conclude that within the parameter ranges considered in this work, the inclusion of optomechanical coupling renders the two-dimensional cavity-magnon system with YIG sphere nonlinearity more favorable for the stabilization of the Mott insulating phase, while photon Kerr nonlinearity promotes the superfluid phase. Our theoretical results provide a series of rich phase diagrams for the superfluid--Mott insulator transition. These behaviors differ from those of the JCH model~\cite{16}, and the distinctive features we uncover may open new possibilities for further studies.

\section{The ergodic to MBL phase transition}\label{IV}

In the previous section, we discussed the superfluid–Mott insulator transition in the cavity optomagnonic system, driven primarily by the competition between particle–particle interactions and kinetic energy. The system undergoes a transition from a long-range coherent superfluid phase to a localized Mott insulating phase. In this section, we explore another localized phase in the system—the MBL phase.

A superfluid phase does not necessarily imply ergodicity~\cite{41}. In the presence of disorder, even when the system exhibits macroscopic coherence, its long-time dynamics may fail to explore the entire Hilbert space, preventing it from reaching thermal equilibrium. This indicates that coherence and ergodicity are distinct concepts~\cite{49}. In particular, for isolated quantum systems, the introduction of disorder can break ergodicity and drive the system into the MBL phase. The MBL phase is a non-equilibrium, non-thermal state with long-term memory, which arises from the interplay of interactions and disorder, contrasting with the purely interaction-driven Mott insulating phase.

Here, we investigate the ergodic–MBL transition in the two-dimensional cavity optomagnonic system. Although both the MBL phase and the previously discussed Mott insulator phase are localized states, the ergodic–MBL transition is a disorder-induced transition~\cite{54}. By tuning the strength of disorder, we induce this phase transition and characterize the system’s behavior from the perspective of ergodicity. As a complement to the coherence-focused study of the superfluid–Mott transition, this section highlights the system’s physics along the dimension of ergodicity.

Extensive previous studies~\cite{44,45,46,47} have shown that various interacting quantum systems—such as the JCH model, the Heisenberg model, and the XXZ spin chain—undergo an ergodic–MBL transition as disorder strength increases, though the critical disorder strength varies with the model. For instance, the MBL transition in a spin-$1/2$ XXZ chain with random fields depends nontrivially on both disorder strength and interaction strength, and is typically characterized by tuning disorder strength at fixed interaction~\cite{46}. The JCH model in the MBL phase displays nonthermalization
behavior~\cite{45}. In this work, we probe the disorder-induced transition by varying the disorder strength in our system.

To analyze the ergodic–MBL transition, we start with the clean system Hamiltonian and introduce the disorder;
\begin{equation}
	H^{\mathrm{clean}} = \sum_{i} H^{cop} - \kappa \sum_{\langle i,j \rangle} a^{\dagger}_{i} a_{j} + K \sum_i m_i^\dagger m_i m_i^\dagger m_i,
\end{equation}

\begin{equation}
	H^{\mathrm{disorder}} = \sum_{i} H^{cop} - \kappa \sum_{\langle i,j \rangle} a^{\dagger}_{i} a_{j} + \sum_i \mathrm{h}_i m_i^\dagger m_i m_i^\dagger m_i,
\end{equation}
where each $\mathrm{h}_i$ is a random variable uniformly drawn from the interval $[-\mathrm{h}_{\max}, \mathrm{h}_{\max}]$. Here $\mu$ should be regarded as a tunable local potential parameter rather than a thermodynamic chemical potential~\cite{51}.
 We characterize the ergodic–MBL transition by calculating the ratio of consecutive level spacings $\langle r \rangle$ and the level spacing distribution. The ratio is defined as
\begin{equation}
	\langle r \rangle = \frac{\min(\delta^n, \delta^{n+1})}{\max(\delta^n, \delta^{n+1})}, \quad \delta^n = E_n - E_{n-1},
\end{equation}
where $E_n$ is the $n$th eigenenergy. To reduce boundary effects, we perform statistics on the middle one-third of the spectrum. This method has been widely used in previous studies~\cite{48,52,54,57}. If the level spacing distribution approaches the Wigner–Dyson distribution,
$P_{\mathrm{WD}}(s) = \frac{\pi s}{2} e^{-\frac{\pi s^2}{4}}$, with $\langle r_{\mathrm{WD}} \rangle \approx 0.531$, the system is in a thermalized (ergodic) phase with level repulsion. Conversely, if it approaches the Poisson distribution, $P_{\mathrm{P}}(s) = e^{-s}$, with $\langle r_{\mathrm{P}} \rangle \approx 0.386$, the system is in the MBL phase, allowing level crossings and nearly independent energy levels.

To visualize the evolution more intuitively, we construct the
disorder–energy density phase diagram $(\mathrm{h},\epsilon)$ using the participation entropy $S^{P}$ as an indicator. The participation entropy $S^{P}$, defined for any eigenstate $|n \rangle$ by
$S^{P}(|n \rangle) = -\sum_{i} \lambda_i \ln \lambda_i, \quad 
\lambda_i = |\langle n | i \rangle|^2 ,$ where $\lambda_i$ is the probability of finding the eigenstate $|n \rangle$ in basis state $|i\rangle$~\cite{48}. In the MBL phase, the participation entropy obeys an area law ($S^{P} \ll 1$), whereas in the ergodic phase it follows a volume law ($S^{P} \simeq 1$). Using these measures, we observe the ergodic-MBL transition in the two-dimensional cavity–optomagnonic system.
As disorder in the nonlinear YIG spheres increases, the system transitions from the ergodic to the MBL phase, marked by a change in $S^{P}$.

It is worth noting that for our two-dimensional cavity optomagnonic system, the Hilbert space dimension grows very rapidly with system size, reaching millions of dimensions for a $4 \times 4$ lattice, making exact diagonalization infeasible. Thus, we focus on a $3 \times 3$ lattice with total excitation number $N = \lfloor \frac{3 \times 3}{2} \rfloor$ (where $\lfloor \cdot \rfloor$ denotes the floor function).

\begin{figure}[!htbp]
	\centering
	\begin{subfigure}[b]{0.3\textwidth}
		\includegraphics[width=\textwidth]{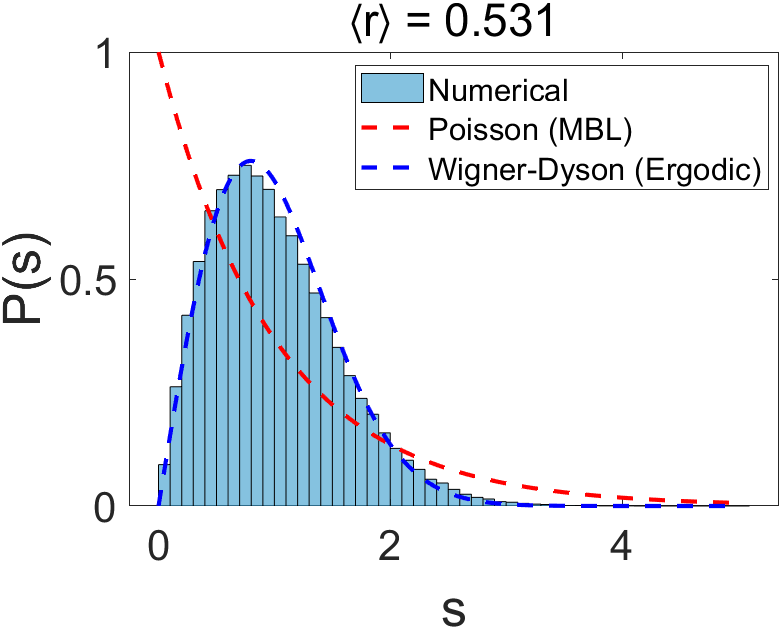}
		\caption{}
		\label{fig:sub1}
	\end{subfigure}
	\hspace{0.01\textwidth}
	\begin{subfigure}[b]{0.3\textwidth}
		\includegraphics[width=\textwidth]{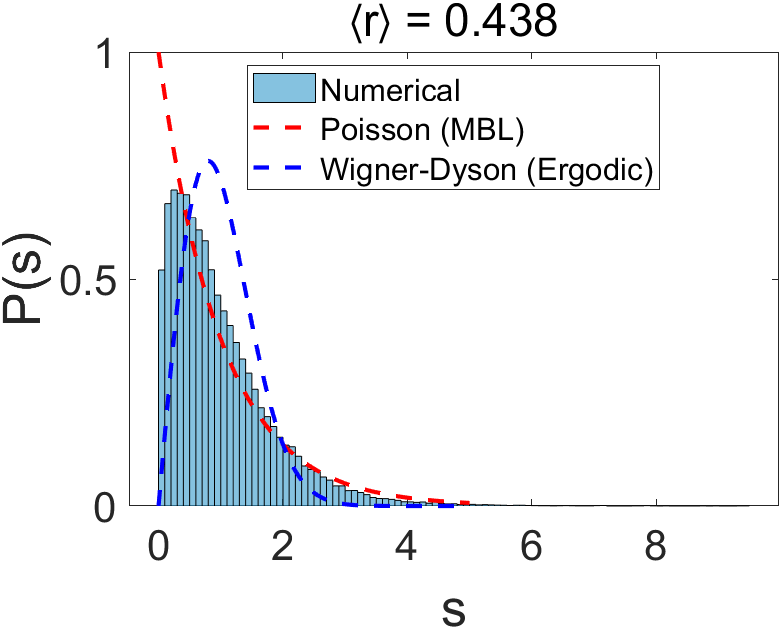}
		\caption{}
		\label{fig:sub2}
	\end{subfigure}
	\begin{subfigure}[b]{0.3\textwidth}
		\includegraphics[width=\textwidth]{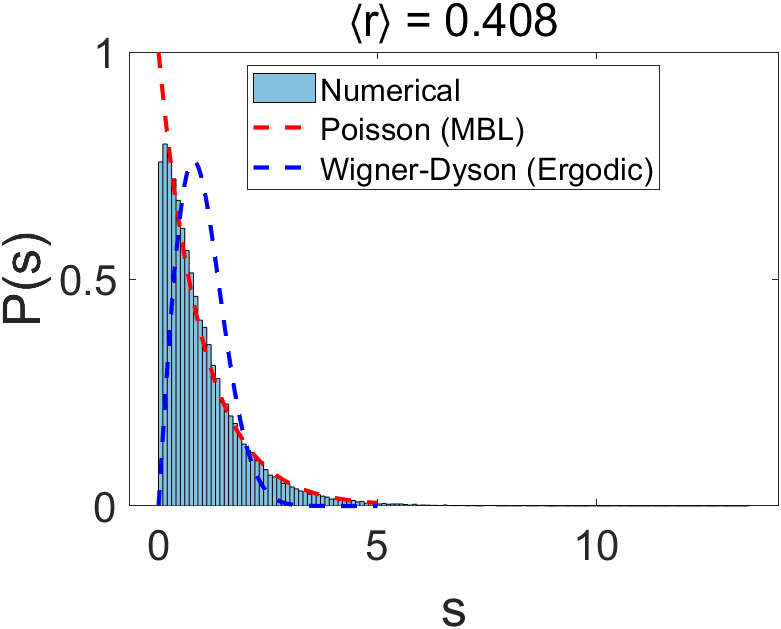}
		\caption{}
		\label{fig:sub3}
	\end{subfigure}
	\\
	\captionsetup{justification=raggedright, singlelinecheck=false}
	\caption{Level spacing ratio $\langle r \rangle$ and level spacing distribution with increasing disorder strength $\frac{h}{g_1}$, \\other parameters are fixed.\quad From (a) to (c), dis-\\order increases.}
	\label{fig:figure6}
\end{figure}

Fig~\ref{fig:figure6} shows the Wigner–Dyson distribution (red) and Poisson distribution (blue) for comparison. Panels (a), (b), and (c) correspond to increasing disorder strengths. At weak disorder (a), the level spacing distribution closely matches Wigner–Dyson, indicating an ergodic phase. At strong disorder (c), the distribution approaches Poisson, indicating MBL. This confirms that the system transitions from ergodic to localized as disorder grows. The behavior of $\langle r \rangle$ corroborates this conclusion. A key advantage of using $\langle r \rangle$ is that it does not require unfolding the spectrum, making it better suited for experimental comparisons. Moreover, previous numerical works~\cite{52,53,54} show $\langle r \rangle$ yields more accurate characterization than level spacing distributions alone.

To complement these results, we compute the participation entropy to construct the phase diagram, clearly showing the regions occupied by the ergodic and MBL phases and the mobility edge separating them.

\begin{figure}[!htbp]
	\centering
	\begin{subfigure}[b]{0.23\textwidth}
		\includegraphics[width=\linewidth]{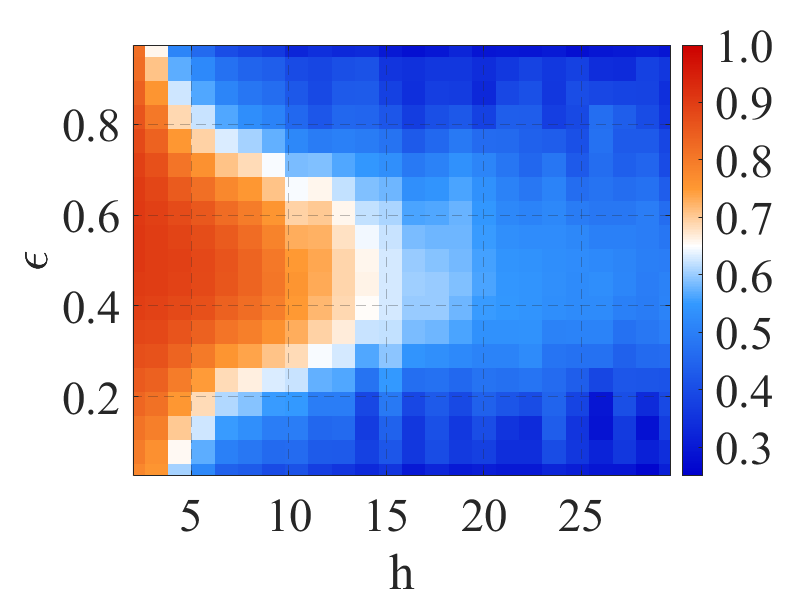}
		\caption{}
		\label{fig:sub1}
	\end{subfigure}
	\hfill
	\begin{subfigure}[b]{0.23\textwidth}
		\includegraphics[width=\linewidth]{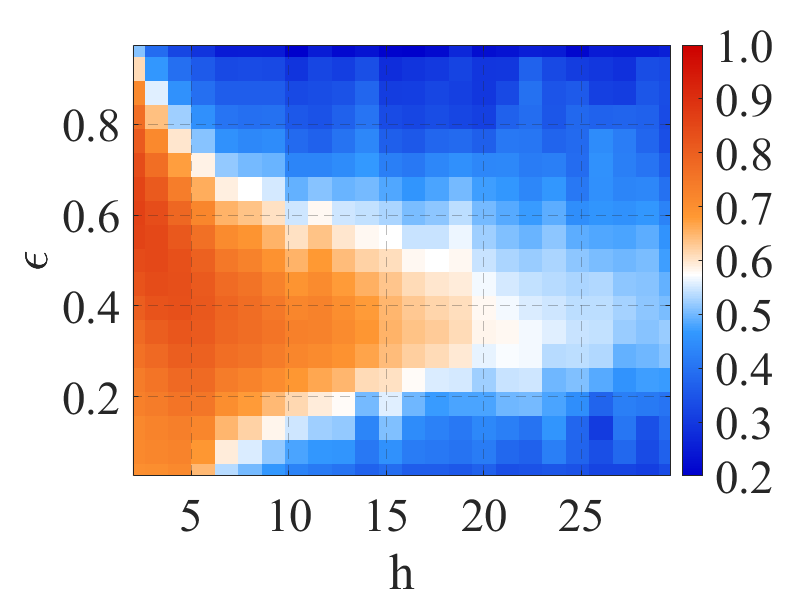}
		\caption{}
		\label{fig:sub2}
	\end{subfigure}
	\captionsetup{justification=raggedright, singlelinecheck=false}
	\caption{Phase diagram in disorder strength $(\mathrm{h})$ vs. energy density $(\epsilon)$ plane (with $g_1$ set to 1). Orange regions correspond to the ergodic phase $(S \simeq 1)$, b-\\lue to the MBL phase $(S \ll 1)$, and the intermedia-\\te region marks the mobility edge. (a) $\frac{\mu}{g_1} = 1$, (b) $\frac{\mu}{g_1} = 5$.}
	\label{fig:figure7}
\end{figure}

Fig~\ref{fig:figure7} illustrates the distribution of ergodic and MBL phases in the $(\mathrm{h}, \epsilon)$ plane. The color scale represents participation entropy, a standard indicator originally used in single-particle localization~\cite{55} and later extended to many-body physics~\cite{48}. The orange regions satisfy a volume law ($S \simeq 1$) characteristic of ergodicity, while blue regions satisfy an area law ($S \ll 1$), indicating localization. The mobility edge separating these regions becomes more pronounced as disorder increases. Comparing panels (a) and (b), increasing chemical potential from $\frac{\mu}{g_1} = 1$ to $\frac{\mu}{g_1} = 5$ enlarges the ergodic region and shifts the phase boundary.

In summary, we have studied the ergodic properties of the two-dimensional cavity optomagnonic system. Increasing nonlinear disorder in the YIG spheres drives a transition from extended ergodic states to localized MBL states. Furthermore, increasing chemical potential promotes ergodicity. Our results from both level statistics and participation entropy are mutually consistent.
Our theoretical study reveals intriguing physical effects associated with the ergodic–MBL transition. Compared with the JCH model~\cite{45}, we introduce additional degrees of freedom, thereby offering a reliable platform for future explorations. This may open up new possibilities for fundamental research on novel quantum phases as well as potential quantum applications.

\section{Conclusions}\label{V}

In conclusion, we have investigated two distinct phase transitions in a two-dimensional cavity optomagnonic system: (i) the superfluid–Mott insulator transition in the clean limit, and (ii) the ergodic–many-body localization transition driven by nonlinear disorder in the YIG spheres.

For the superfluid–Mott insulator transition, we find that introducing phonon–photon coupling enhances the coherence of the system in the presence of Kerr nonlinearity from the YIG spheres. In contrast, the photonic Kerr nonlinearity tends to enhance the Mott insulating phase more effectively than its magnonic counterpart. To further clarify the physical mechanism, we computed the effective repulsive potential under photonic Kerr nonlinearity.
Regarding the ergodic–MBL transition, our results show that increasing the disorder strength in the Kerr nonlinearity drives the system from the ergodic phase into the MBL phase. Conversely, increasing the chemical potential expands the ergodic region. These findings provide a clear characterization of the phase transitions and underlying mechanisms in the two-dimensional cavity optomagnonic system.                                                                                                                                                                                                                                                                                                                                                                                                                                                                                                                                                                                                                                                                                                                                            

By investigating both the superfluid–Mott and ergodic–MBL transitions within a single physical platform, our study showcases the promising potential of cavity optomagnonic systems for exploring complex many-body quantum phenomena. This opens avenues for future experimental and theoretical investigations of nonequilibrium quantum phase transitions, thermalization dynamics, and disorder effects in hybrid quantum systems. Additionally, this work provides an interesting platform for simulating quantum states in cavity optomagnonic systems.

Experimentally, several experiments have successfully realized magnon–photon strong-coupling systems~\cite{PhysRevLett.113.083603,PhysRevLett.111.127003,PhysRevLett.120.057202,zhang_observation_2017-2}. In particular, the magnon Kerr effect has been demonstrated in a cavity–magnon setup using a YIG sphere~\cite{PhysRevB.94.224410}, pronounced photonic Kerr nonlinearity has been observed in an optical cavity~\cite{PhysRevLett.132.143602}, and nonclassical phonon–photon correlations have been achieved in a hybrid optomechanical resonator~\cite{PhysRevLett.133.206401}. Additionally, recent theoretical studies have explored the cross-Kerr nonlinearity between two magnon modes in antiferromagnetic insulators, which can be large in materials with weak easy-axis magnetic anisotropy, potentially enabling magnon and photon blockade effects in cavity systems~\cite{m2rq-l6fl}. Building upon these developments, the cavity optomagnonic system proposed in this work is promising for near-future experimental realization.

\section*{ACKNOWLEDGMENTS}
This work was supported by National Natural Science Foundation of China(Grants No. 11874190 and No. 12247101). Support was also provided by Supercomputing Center of Lanzhou University.

\bibliography{REF}

\end{document}